# Deep Learning Based Apparent Diffusion Coefficient Map Generation from Multi-parametric MR Images for Patients with Diffuse Gliomas

Zach Eidex[1], Mojtaba Safari[1], Jacob Wynne[1], Richard L.J. Qiu[1], Tonghe Wang[3], David Viar Hernandez[1], Hui-Kuo Shu[1,5], Hui Mao[4,5] and Xiaofeng Yang[1,2,5*]

[1]Department of Radiation Oncology, Emory University, Atlanta, GA

[2]School of Mechanical Engineering, Georgia Institute of Technology, Atlanta, GA

[3]Department of Medical Physics, Memorial Sloan Kettering Cancer Center, New York, NY

[4]Department of Radiology and Imaging Sciences, Emory University, Atlanta, GA

[5]Winship Cancer Institute, Emory University, Atlanta, GA

**Running title:** ADC Map Synthesis

**Manuscript Type:** Original Research

**Contact information:**
Email – xiaofeng.yang@emory.edu
Address - 1365-C Clifton Road NE Atlanta, Georgia 30322

# ABSTRACT

**Purpose:** Apparent diffusion coefficient (ADC) maps derived from diffusion weighted magnetic resonance imaging (DWI MRI) provides functional measurements about the water molecules in tissues. However, DWI is time consuming and very susceptible to image artifacts, leading to inaccurate ADC measurements. This study aims to develop a deep learning framework to synthesize ADC maps from multi-parametric MR images.

**Methods:** We proposed the multiparametric residual vision transformer model (MPR-ViT) that leverages the long-range context of vision transformer (ViT) layers along with the precision of convolutional operators. Residual blocks throughout the network significantly increasing the representational power of the model. The MPR-ViT model was applied to T1w and T2- fluid attenuated inversion recovery images of 501 glioma cases from a publicly available dataset including preprocessed ADC maps. Selected patients were divided into training (N=400), validation (N=50) and test (N=51) sets, respectively. Using the preprocessed ADC maps as ground truth, model performance was evaluated and compared against the Vision Convolutional Transformer (VCT) and residual vision transformer (ResViT) models with the peak signal-to-noise ratio (PSNR), structural similarity index measure (SSIM), and mean squared error (MSE).

**Results:** The results are as follows using T1w + T2-FLAIR MRI as inputs: MPR-ViT - PSNR: $31.0 \pm 2.1$, MSE: $0.009 \pm 0.0005$, SSIM: $0.950 \pm 0.015$. In addition, ablation studies showed the relative impact on performance of each input sequence. Both qualitative and quantitative results indicate that the proposed MR-ViT model performs favorably against the ground truth data.

**Conclusion:** We show that high-quality ADC maps can be synthesized from structural MRI using a MPR-VCT model. Our predicted images show better conformality to the ground truth volume than ResViT and VCT predictions. These high-quality synthetic ADC maps would be particularly useful for disease diagnosis and intervention, especially when ADC maps have artifacts or are unavailable.

## INTRODUCTION

Gliomas, graded in severity from I to IV by the World Health Organization (WHO), represent over half of malignant tumors that affect the central nervous system.[1] Low grade gliomas (LGGs; WHO grades I and II) are less aggressive and offer a significantly improved prognosis compared to high grade gliomas (HGGs; WHO grades III and IV). The WHO grade is determined, along with molecular and genomic biomarkers by visualizing the tumor progression through medical imaging.[2] Given magnetic resonance imaging's (MRI) excellent soft tissue contrast, MRI has become a modality of choice to diagnose and prognosis for diffuse glioma.[3] MRI sequences including T1-weighted (T1w), T1-weighted postcontrast (T1c), T2-weighted (T2w), and T2 fluid attenuated inversion recovery (FLAIR) MRI provide precise tumor localization and structural information like peritumoral edema, necrosis, and the mass effect.[4] However, tumor differentiation can still be a challenge with structural MRI since glioma grades can have similar appearances, so advanced techniques that capture the functional information of the tumor are often needed.

Diffusion weighted MRI (DWI) measures changes in cellular water mobility. Since tumors are typically marked by an increase in cellularity and, therefore, reduced water mobility, these differences in water diffusivity can be quantified by taking multiple DWIs at differing gradient strengths to create an apparent diffusion coefficient (ADC) map. DWI and ADC maps have demonstrated promise in determining glioma prognosis.[5] However, DWI and ADC are highly susceptible to image artifacts due to the fast imaging acquisition techniques that are inherently susceptible to being corrupted by image artifacts.[6] To address this issue, previous studies used deep learning models to generate ADC maps from under-sampled DWI to reduce acquisition time and the likelihood of artifacts.[7,8] Another approach removes the need for DWI altogether by leveraging structural MRI to predict the ADC map.[9,10] Unique to this study, we find that using multiple MR sequences (T1w and T2-FLAIR) to generate the synthetic ADC map volume allows for higher quality predictions than a single sequence alone.

While there are relatively few studies specific to ADC map synthesis, Intramodal MRI synthesis has seen significant progress alongside advances in advancements in natural image translation techniques. Most approaches use Pix2pix which claims to produce more realistic results by using generative adversarial network (GAN) with U-Net as a generator.[11,12] While Pix2pix can generate reasonable predictions, Pix2pix is exclusively based on convolutional layers so can struggle to capture long-range context. Recent approaches implement vision transformers (ViT) that capture long-range context by finding relationships throughout the entire feature map.[13,14] To mitigate the computational burden of vision transformers, hybrid CNN-transformers strategically place ViT layers in deeper, more abstract layers where feature map sizes are smaller and the attention calculation is less costly. The residual vision transformer (ResViT) model incorporates both vision transformers and a GAN architecture and supports multimodal inputs.[15] However, the GAN architecture was found to introduce considerable training complexity and was not helpful in generating high-quality ADC maps. In addition, ResViT called for initially pretraining the convolutional layers first without transformer layers and pretraining the transformer layers on natural images but were not found to improve performance. By using only the ResViT generator and implementing several efficiency improvements like flash attention and the ability to natively use any resolution with dimensions divisible by eight, the vision convolutional neural network transformer (VCT) model is proposed for structural MRI to ADC map synthesis.[16-18]

In this study, we build off the VCT model and propose the multiparametric residual vision transformer (MPR-ViT) model to produce highly accurate ADC maps. We make the following contributions:

(1) This is the first attempt to produce synthetic ADC maps for patients with diffuse glioma or tumor volumes. We evaluate the model's performance on heterogeneous regions compared to other state-of-the-art methods.
(2) Leveraging efficiency improvements made in the VCT model, we are more easily able to significantly increase the VCT model's representational power by replacing each encoder and decoder convolutional block with 3 residual blocks.
(3) We achieve state-of-the-art performance for single-modal (T1w → ADC map and T2-FLAIR → ADC map) along with multimodal (T1w + T2-FLAIR → ADC map) image synthesis with the MPR-ViT model.

1. **METHOD**

**2.1 Data Acquisition and Preprocessing**

Patients were selected from the publicly available The Cancer Imaging Archive (TCIA) University of California San Francisco Preoperative Diffuse Glioma (UCSF-PDGM) containing 501 adult patients with T1w and T2-FLAIR MRI along with ADC maps.[19,20] All scans were performed on a 3.0T scanner (Discovery 750, GE Healthcare, Waukesha, Wisconsin, USA) and dedicated 8-channel head coil (Invivo, Gainesville, Florida, USA) between 2015 and 2021. The patients all had histopathologically confirmed grade II-IV diffuse gliomas including 55 (11%) grade II, 42 (9%) grade III, and 403 (80%) grade IV tumors. Segmentation maps were created from an ensemble model using top ranking segmentation algorithms for the whole brain volume as well as the three major tumor regions: enhancing tumor, necrotic tumor, and peritumoral edema.[21] These segmentation maps were then corrected by trained radiologists and approved by 2 expert reviewers. The ADC maps were produced from 2D 55-direction high angular resolution diffusion imaging and were corrected for Eddy currents. These DWI were then registered and resampled to the T2-FLAIR image at 1 mm isotropic resolution through automated non-linear registration, and skull stripped with a publicly available deep-learning algorithm.[22,23] These images were then transformed into ADC maps. All volumes downsampled by a factor of 2 using bicubic interpolation. The down-sampling was done to reduce computational burden.

**2.2 MPR-ViT Architecture**

Figure 1 illustrates the proposed architecture, which T1-weighted (T1w) and 3T T2-FLAIR axial slices are concatenated channel-wise to produce synthetic ADC maps. The encoder and decoder learn localized features through convolutional layers connected with residual skip connections while the information bottleneck encourages learning more abstract, long-range features though residual convolutional and transformer layers. The encoder and decoder each consist of 3 combined residual blocks defined to be three sequential residual blocks of the same dimensionality. Compared with single residual blocks, the combined residual blocks allow for increased abstraction and representational power. The information bottleneck similarly allows for a high degree of abstraction by incorporating 11 layers connected with residual skip connections and does not change the dimensions of the feature map. Two of these layers incorporate transformers to allow for long-range context and share weights to decrease the computational burden. In total, MPR-ViT contains 27 residual blocks and 2 ViT blocks.

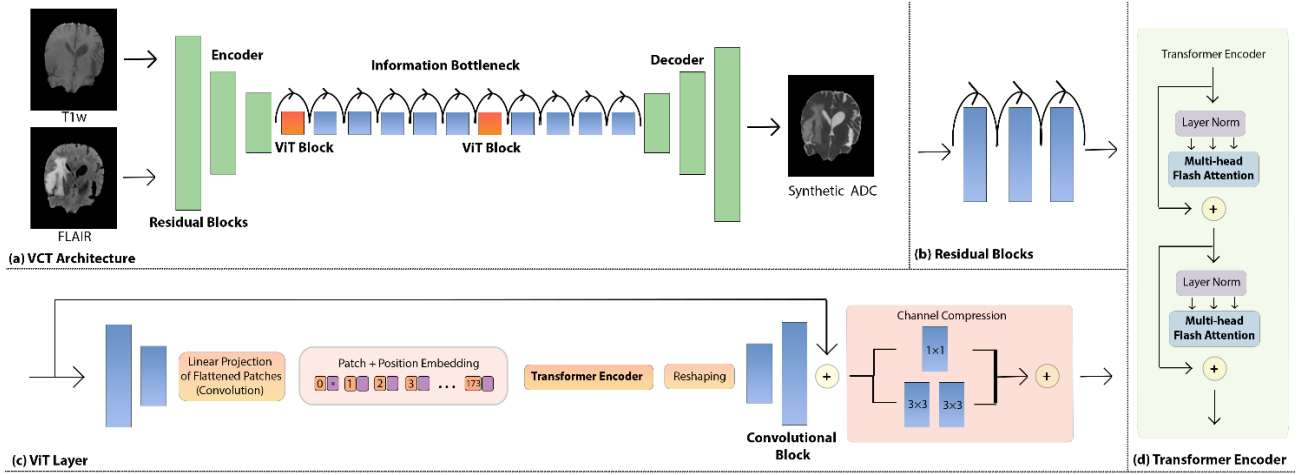

**Figure 1.** (a) Schematic flow chart of the MPR-ViT model. The green, orange, and blue blocks are combined residual, ViT, and convolutional blocks, respectively. Curled arrows represent residual skip connections bypassing the in-between layer to facilitate training deeper networks. (b) Each combined residual block shown in (a) corresponds to 3 residual blocks. These additional layers increase the depth and representational capacity of the model. (c) The feature maps are downsampled with convolutional blocks, flattened, and coupled with position embeddings before being passed into the transformer encoder. After upsampling, a residual skip connection prevents the loss of previously derived context. Finally, the channel compression module better reduces the number of feature maps than simple downsampling. (d) A standard transformer encoder with the inclusion of the flash attention mechanism.

### 2.3 Encoder and Decoder

The encoder and decoder are each composed of 3 combined residual blocks. Each combined residual block contains 3 residual blocks containing a convolutional layer followed by a norm layer and rectified linear unit (ReLU) activation function. All convolutional layers have a kernel size of 3×3 except the very first and last layers which have a large kernel size of 7×7 to capture broader the receptive field. Each convolutional block has a residual skip connection to facilitate the deeper network by backpropagating errors. The final convolutional layer of the combined residual block in the encoder has a stride of 2 to reduce the dimensionality by a factor of 2 while the combined residual block of the decoder is a transposed convolution which doubles the feature map size. In total the dimensionality is reduced by a factor of 4 to a resolution of 30×30 before being passed into the information bottleneck and ultimately returned to the original input size of 120x120 in encoder. This encourages the information bottleneck to focus on course details and reduces the computational burden of the ViT blocks. The encoder and decoder are symmetric except that the final convolutional block in the decoder sets the output number of channels to 1 compared to the 2 input channels in the encoder. In addition, the final layer in the network was followed by a hyperbolic tangent activation function to ensure the range of the output feature map values is from -1 to 1.

### 2.4 Information Bottleneck

The information bottleneck captures abstract, global context using both convolutional blocks and powerful but computationally expensive ViT blocks. Residual skip connections are placed across both blocks to mitigate the vanishing gradient problem by creating an alternate, shorter path for the gradient during backpropagation.[24] The convolutional blocks in the information bottleneck shown in Figure 1 are comprised of two sequential 3 × 3 convolutional layers which are then connected to the input feature map through a skip connection. The ViT blocks similarly incorporate two sequential 3 × 3 convolutional layers but use a stride of 2 which reduces the dimensionality of the input feature maps by a factor of 4 to ease the computational burden. Afterwards, the feature maps are flattened and given each feature on the feature map is given before being input into the transformer layer. After the transformer layer, two 3 × 3 transposed convolutional layers return the feature maps to their original spatial dimensions before passing through the ViT block. Finally, this output is connected to the input feature map through a residual skip connection. Since the ViT block downsamples the feature map from 30×30 by a factor of 4 rounding to 8×8, then upsampling to 32x32, we implement bilinear interpolation to resize it back to 30x30. We chose this approach in favor of padding the input images by 128×128 to improve efficiency and because we did not notice a decline in performance. Finally, we replace the traditional attention mechanism with more efficient FlashAttention.[25] Read-and-write operations to GPU memory are the primary bottleneck for the attention mechanism's calculation speed in transformers. Flash Attention addresses this issue by optimizing the use of the GPU's small, high-performance SRAM cache through tiling and recomputation. This approach lessens the memory demand while also achieving a notable increase in speed and remains computationally equivalent to the traditional attention mechanism.

## 2.5 Implementation Details

The MPR-ViT model was trained on a consumer-grade NVIDIA RTX 4090 GPU with 24 GB of memory, and additional results were gathered with a cloud-based NVIDIA A10 with 24 GB of memory. The dataset was augmented by randomly flipping the images in the coronal plane. An AdamW gradient optimizer (learning rate 2e-4, β1 = .500, β2 = .999, eps = 1e-6) was set to optimize the learnable parameters over 251 epochs or when the model no longer reduced the validation loss. For the multimodal hold-out test, training was stopped at 171 epochs. The AdamW optimizer was chosen to minimize the loss function (L1 loss) for its improved generalization performance over the Adam optimizer due to a decoupling of the weight decay and gradient update.[26] 32 image slices were used for each batch. Each epoch took approximately 4 minutes, and the inference time of each image slice during testing was 3.4 milliseconds (ms) or 238 ms for a typical patient volume with 70 slices.

## 2.6 Validation and Evaluation

Model performance was assessed using a hold-out test by randomly dividing a total of 501 patients into training (400 patients: 29,005 slices), validation (50 patients: 3,511 slices), and testing (51 patients: 3,590 slices) slices. Results were quantified using mean squared error (MSE), peak signal-to-noise ratio (PSNR), and structural similarity index (SSIM) over the entire 3D volume. Student's two-sided t-test was employed to compare the VCT model results with comparative methods. The significance level was set at 0.05 for these evaluation metrics. MSE measures the voxel wise difference between the synthetic and ground truth volumes such that a value of zero means no difference.[27] PSNR is inversely related with the MSE so that higher PSNR values correspond to higher similarity to the ground truth volume. Logarithmic scaling was applied to make it more closely align with human perception.[28] SSIM considers luminance, contrast, and structural similarity functions to most closely align with human perception. SSIM values range from -1 to 1 with 1 being perfect correspondence with the ground truth volume.[29] MSE, SSIM, and PSNR are defined below *where n is the total number of voxels, $X_i$ and $Y_i$ are the voxel intensity of the synthetic and ground truth volumes, and $MAX_I$ is the maximum possible voxel value of the ground truth volumes.*

$$MSE = \frac{1}{n} \sum_{i=1}^{n}(X_i - Y_i)^2 \qquad (1)$$

$$PSNR = 10 \, log\left(\frac{MAX_I^2}{MSE}\right) \qquad (2)$$

$$SSIM = l(x,y) \cdot c(x,y) \cdot s(x,y) = \frac{(2\mu_x\mu_y + C_1)(2\sigma_{xy} + C_2)}{(\mu_x^2 + \mu_y^2 + C_1)(\sigma_x^2 + \sigma_x^2 + C_2)} \tag{3}$$

The SSIM is comprised of luminescence ($l(x,y)$), contrast ($c(x,y)$), and structural similarity ($s(x,y)$) functions. $\mu_x$ and $\mu_y$ are the means of the synthetic and ground truth volumes, $\sigma_x^2$, $\sigma_y^2$, and $\sigma_{xy}$ are the variance of the synthetic volume, the variance of the ground truth volume, and the covariance between the synthetic and ground truth ADC maps respectively. C₁ and C₂ are small constants to avoid division by zero; $C_1 = 0.01L$ and $C_2 = 0.02L$ where L osthe maximum value of the target.

To assess the complexity of the model, the number of parameters and floating-point operations per second (FLOPS) of the model were calculated.

## 2. RESULTS

Synthetic ADC maps generated by the MPR-ViT model are compared against the ground truth ADC and evaluated through the MSE, PSNR, and SSIM metrics (see Table 1). Compared with the ResViT and VCT architectures, MPR-ViT showed a statistically significant improvement (p-value < .001) using both unimodal and multimodal inputs achieving an MSE: 0.0009 ± 0.0005, PSNR: 30.9 ± 2.0, SSIM:0.950 ± 0.015. Both VCT and MPR-ViT benefited from a multimodal approach while ResViT achieved its highest performance with T2-FLAIR only inputs. Example cases of output images from 4 patients are shown in Figure 2 with zoom-in on regions along with ground truth ADC maps and T1w and T2-FLAIR MRI input images. Figure 2a-c show patients with the tumor volumes while Figure 2d shows a tumor-free region. VCT most closely resembles the ground truth (GT) ADC maps while ResViT produces the worst outputs.

Configurations of the VCT model with increasing numbers of residual blocks added to the combined residual blocks are evaluated in Table 2. The VCT model saw marked improvements in the SSIM value with the addition of one and two residual blocks but gains in performance after 3 layers. Model complexity rose predictably in terms of model size and computational cost with additional residual blocks leading to longer training times.

**Table 1**. Quantitative results for the whole brain volume. P-value metrics are compared against MPR-ViT's performance. The arrows indicate the direction of the better metric value.

| | | MSE[↓] | PSNR[↑] | SSIM[↑] | P-Value |
|---|---|---|---|---|---|
| **T1w Only** | | | | | |
| | T1w MRI | .0105 ± .0036 | 20.0 ± 1.5 | .692 ± .040 | <.001 |
| | ResViT | .0014 ± .0005 | 28.8 ± 1.5 | .925 ± .019 | <.001 |
| | VCT | .0012 ± .0004 | 29.6 ± 1.6 | .934 ± .017 | <.001 |
| | MPR-ViT | **.0011 ± .0005** | **30.1 ± 1.9** | **.939 ± .017** | X |
| **FLAIR Only** | | | | | |
| | FLAIR MRI | .0144 ± .0091 | 19.4 ± 3.0 | .768 ± .034 | <.001 |
| | ResViT | .0013 ± .0005 | 29.1 ± 1.5 | .932 ± .016 | <.001 |
| | VCT | .0022 ± .0008 | 26.9 ± 1.5 | .889 ± .025 | <.001 |
| | MPR-ViT | **.0010 ± .0005** | **30.4 ± 1.9** | **.945 ± .015** | X |
| **T1w + FLAIR** | | | | | |
| | ResViT | .0014 ± .0005 | 28.9 ± 1.5 | .927 ± .017 | <.001 |
| | VCT | .0010 ± .0006 | 30.4 ± 2.1 | .945 ± .016 | <.001 |
| | MPR-ViT | **.0009 ± .0005** | **31.0 ± 2.1** | **.950 ± .015** | X |

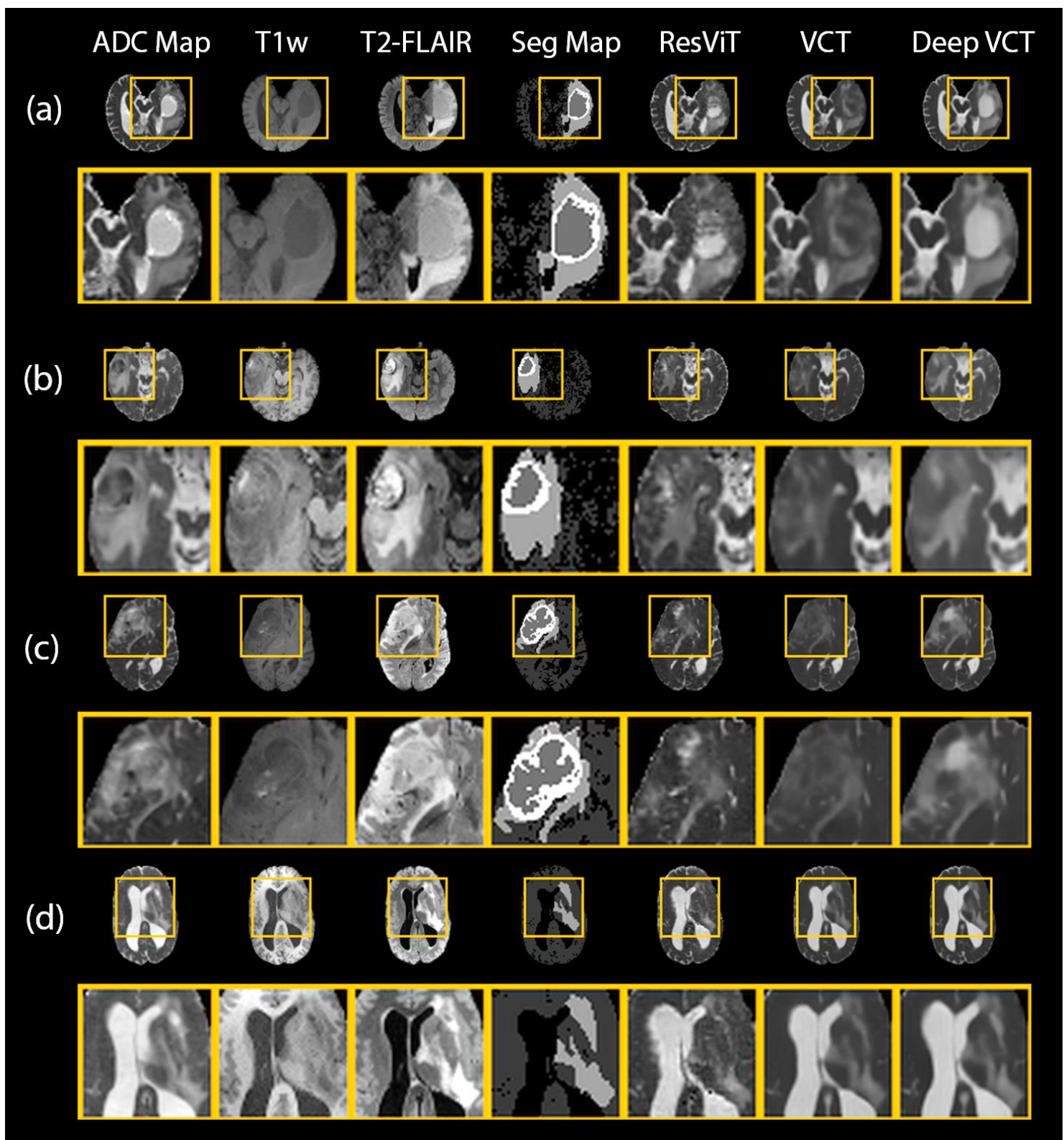

**Figure 2.** Example ADC maps generated from ResViT, VCT, and MPR-ViT from 4 patients along with zoom-in on interesting regions. All methods took both T1w and T2-FLAIR MRI as inputs. (a) - (d) show reconstructions of slices containing tumor. Shown in the segmentation map in order from darkest to lightest are the necrotic tumor core, brain mask, peritumoral edema, and enhancing tumor.

**Table 2.** Model Size and Performance. The VCT model improved by adding additional residual blocks to each combined residual block up to 2 residual blocks (RB). Since there are 6 combined ~~residual blocks~~RBs in the network, VCT + 1 RB corresponds to adding 6 RB total. All models were trained on T1w + T2-FLAIR and with a batch size of 32.

| | Model Size (Millions of Parameters) | FLOPS (Billions) | SSIM |
|---|---|---|---|
| ResViT (Generator) | 123 | 1378 | .927 ± .017 |
| ResViT (Discriminator) | 4 | 33 | X |
| VCT | 115 | 877 | .945 ± .016 |
| VCT + 1 RB | 117 | 1068 | .949 ± .016 |
| VCT + 2 RB (MPR-ViT) | 119 | 1277 | .951 ± .016 |
| VCT + 3 RB | 121 | 1486 | .948 ± .015 |
| VCT + 4 RB | 123 | 1694 | .950 ± .015 |

## 3. DISCUSSION

In this study, we propose the MPR-ViT model for multimodal ADC map synthesis which outperforms the ResViT and VCT models as measured by MSE, PSNR, and SSIM and based on qualitative results. By generating accurate ADC maps from multi-modal, structural MRI, greater contrast and detail can be achieved when DWI contains artifacts or is unavailable. The synthetic ADC maps generated by the MPR-ViT model are, overall, highly conformal to the ground truth volume even in difficult heterogeneous and tumor regions. These favorable image characteristic may have an increased ability for tumor characterization and detection so synthetic ADC maps may prove to be clinically useful by demonstrating superior clinical information.[30,31]

To our knowledge, this is the first work to synthesize ADC maps from multimodal MRI. However, several notable studies have been published for multi-modal and ADC map MR image translation tasks.[32] Modifying the StarGAN architecture for intramodal MRI synthesis, Dai et al generate four MR sequences (T1w, T2w, T1-postcontrast (T1c), and T2 Fluid attenuated inversion recovery (FLAIR)) simultaneously from a single input sequence. Using T1w MRI as input, the SSIM for T1c, T2w, and T2-FLAIR were 0.974±0.059, 0.969±0.059, and 0.959±0.059 respectively.[33,34] Wang et al incorporated unsupervised and supervised learning to synthesize ADC maps from T2w MRI with the ultimate goal of localizing prostate cancer lesions, reporting a Fréchet inception distance score of 178.2 ± 3.7.[35,36]

Comparing the input T1w and T2-FLAIR MRI with the synthetic ADC maps, this study reveals that the synthetic ADC maps provide a markedly higher contrast and reveal additional detail, especially of the tumor volume for instance in Figure 2A. In addition, Table 1 demonstrates the usefulness of a multi-modal approach to intramodal MR image synthesis where VCT and MPR-ViT perform the best with T1w + T2-FLAIR MRI as input modalities. We believe this is because the two modalities together provide a richer feature space than a single sequence especially since the peritumoral edema appears as a hyperintense signal on T2-FLAIR.[37]

However, ResViT performed better using only T2-FLAIR as input. In our previous work, we explored the process of generating 7T ADC maps from 3T ADC maps + T1w MRI, we observed instances where ResViT became confused by the T1w MRI. We speculate that the same behavior might be at play in our current work as well.[18] Shown in Figure 2, the synthetic ADC maps generated by the MPR-ViT model are highly similar to the ground truth ADC maps. Moreover, the MPR-ViT model demonstrated the ability to recover detail only visible in ground truth ADC map with especially impressive results in Figure 2a. However, we also note that the synthetic ADC maps are imperfect, especially in heterogeneous regions. This is best shown in the zoomed in region of Figure 2b in which the original tumor volume contains necrosis beneath an edema region The MPR-ViT model achieved the most similar results in this example but still did not capture the necrosis region although it can be seen in the T1w scan. In addition, we note that the ResViT model appears to generate images with less blurring than MPR-ViT, but also adds extraneous details that are not true to the ground truth ADC maps. We believe this is a direct result of ResViT's GAN architecture designed to make images appear more realistic at the cost of giving less importance to pixel-wise accuracy. As verified quantitatively in Table 1, the MPR-ViT model outperforms ResViT and VCT, establishing a new state-of-the-art in all evaluation metrics (MSE, PSNR, and SSIM).

The primary difference between the VCT and MPR-ViT models is the addition of 2 residual blocks forming a combined residual block. As a model is made deeper, it is intuitive that the representational power should increase leading to better performance. For example, the error rate decreases with additional layers in ResNet models with ResNet-152 showing significantly better performance than ResNet-50.[38] However, shown in Table 2, the VCT model did not show additional improvement after 2 additional residual blocks. A potential reason for this is that larger models are more difficult to train and require more data to benefit from the additional parameters. In terms of computational cost, MPR-ViT required approximately 4 million additional parameters and 46% more FLOPS than VCT, but this was still manageable and less than ResViT because of the earlier improvements in the VCT model. Importantly, MPR-ViT was still capable of running on 24 GB of memory with a batch size of 32.

We acknowledge several limitations of the present work. The MPR-ViT model presented here is trained on 2D slices and so does not directly capture the full 3D context of the input data. While the convolutional layers employed in VCT provide an efficient way to capture the fine details, transformers could be added to the encoder and decoder although at a high computational cost due to the larger feature map sizes. Also, from our ablation studies adding residual blocks, we note that increased model complexity can degrade performance, so any additional transformer layers would need to be added carefully. The dataset was comprised of only glioma patients, so it remains to be seen how the results will generalize to patients with other diseases. However, given the successful application of related image translation tasks in these settings, we are optimistic about the generalizability of this model.[39-41] We intend in future work to incorporate full 3D context, train on larger more diverse datasets as they become available[42,43], explore diffusion models[44], and see if these synthetic ADC maps can be used to differentiate LGG from HGG. In addition, we would also like to see if including under-sampled DWI as an input sequence, in addition to structural MRI, further improves the ADC map predictions.

## 5. CONCLUSION

This study presents a deep residual hybrid CNN-transformer model designed for multimodal ADC map synthesis. Our model achieved marked improvements in accuracy compared to current state-of-the art methods were achieved while still being efficient enough to be practical. The proposed method shows great promise in making valuable information in ADC maps more readily available.

## ACKNOWLEDGMENTS

This research is supported in part by the National Cancer Institute of the National Institutes of Health under Award Numbers R01CA272991, R56EB033332 and P30 CA008748. This work was supported in part by Oracle Cloud credits and related resources provided by Oracle for Research.

**Disclosures**

The author declares no conflicts of interest.